\documentclass[letter,scriptaddress,twocolumn, prl,showkeys]{revtex4}
\usepackage{amssymb}
\usepackage{amsmath}
\usepackage{makeidx}
\usepackage{amsfonts}
\usepackage{graphicx,epstopdf}
\usepackage[ansinew]{inputenc}
\usepackage[usenames,dvipsnames]{pstricks}
\usepackage{subfigure}
\usepackage{pdfpages}
\usepackage{epsfig}
\usepackage{pst-grad}
\usepackage{pst-plot}
\usepackage[colorlinks,hyperindex]{hyperref}

\hypersetup{
colorlinks,
citecolor=blue,
linkcolor=black,
urlcolor=black,
}

\makeindex

\begin{document}

\title{Information-Entropic Measure of Energy-Degenerate Kinks in Two-Field
Models}
\author{R. A. C. Correa$^{1,\text{ }2,}$\footnote{E-mail: fis04132@gmail.com}, A. de Souza Dutra$^{1,}$\footnote{E-mail: dutra@gmail.com}, and M.
Gleiser$^{2,}$\footnote{E-mail: mgleiser@dartmouth.edu}}
\affiliation{$^{1}$UNESP-Campus de Guaratinguetá-DFQ, Av. Dr. Ariberto Pereira Cunha, 333
C.P. 205 12516-410 Guaratinguetá SP Brasil\\
$^{2}$Center for Cosmic Origins and Department of Physics and Astronomy,
Dartmouth College, Hanover, NH 03755, USA}

\begin{abstract}
We investigate the existence and properties of kink-like solitons in a class
of models with two interacting scalar fields. In particular, we focus on
models that display both double and single-kink solutions, treatable
analytically using the Bogomol'nyi--Prasad--Sommerfield bound (BPS). Such
models are of interest in applications that include Skyrmions and various
superstring-motivated theories. Exploring a region of parameter space where
the energy for very different spatially-bound configurations is degenerate,
we show that a newly-proposed momentum-space entropic measure called
Configurational Entropy (CE) can distinguish between such energy-degenerate
spatial profiles. This information-theoretic measure of spatial complexity
provides a complementary perspective to situations where strictly
energy-based arguments are inconclusive.
\end{abstract}

\date{August 29, 2014}
\maketitle

\section{1. Introduction}

Since the Scottish channel engineer John Scott Russell first discovered the
existence of solitary waves in 1834 \cite{Russell} and, in particular, since
the 1960s and 70s \cite{Whitham}-\cite{Rajaraman1}, the study of nonlinear
solutions of PDEs that preserve their spatial profile has attracted much
interest in many areas of physics, such as in cosmology \cite{Vilenkin},
field theory \cite{Weinberg, Vachaspati}, condensed matter physics \cite%
{Bishop}, and others \cite{Gu}. In high-energy physics, solitons \cite%
{Weinberg}-\cite{Rajaraman2} are generally known as solutions of nonlinear
field equations whose energy density is localized in space. Certain soliton
solutions, as in the case of sine-Gordon kinks \cite{Rajaraman2}, have the
interesting feature of keeping their shape unaltered after scattering with
other solitons. (Here, we will use \textquotedblleft
soliton\textquotedblright\ to characterize solutions with localized
energy-density, even if many may not maintain their spatial profile after
scattering.)

Nowadays, the properties of nonlinear configurations are well understood in
a wide class of models with or without spontaneous symmetry breaking, and
with or without a nontrivial topological vacuum structure. Of particular
interest to us here are kinks, non-dissipative solutions with an associated
topological charge. Kink configurations arise in (1+1)-dimensional field
theories when the scalar field potential has two or more degenerate minima.
A well-known example is the $\phi ^{4}$-kink, also called the $Z_{2}$ kink 
\cite{Vachaspati, Dashen}. In this case, a single real scalar field $\phi $
interpolates between the two degenerate minima of the potential.

A powerful insight to solve nonlinear problems analytically was introduced
by Bogomol'nyi \cite{Bogomolnyi}, Prasad and Sommerfield \cite{Prasad}.
Known as the BPS bound, it is based on obtaining a first-order differential
equation from the energy functional. With this method, it is possible to
find solutions that minimize the energy of the configuration while ensuring
their stability. A large variety of models in the literature use the BPS
approach, such as solutions found in Skyrme models \cite{Adam, laf},
monopoles \cite{Bobby, Casana}, supersymmetric black holes \cite{Halmagyi},
supergravity \cite{Cassani}, and $K$-field theories \cite{Adam2}.

A few decades ago, it was shown that it is possible to find kink-like
solutions for certain coupled scalar field theories in $(1+1)$-dimensional
models. Presented by Rajaraman, the approach is based on a ``trial and
error'' method which leads to important particular solutions \cite%
{rajaraman3}. Bazeia and collaborators \cite{Bazeia} showed that solutions
of certain second-order differential systems with two or more scalar fields
can be mapped into a corresponding set of first-order nonlinear differential
equations, so that one can obtain the general solution of the system \cite%
{PLBdutra}.

In an apparently disconnected topic, in 1948 Shannon defined the entropy of
a data string as a measure of how much information is needed to characterize
it in a transmission: the more information is needed for a reliable
transmission, the higher the entropy. Inspired by Shannon, Gleiser and
Stamatopoulos (GS) recently proposed a measure of complexity of a localized
mathematical function \cite{PLBgleiser-stamatopoulos}. GS proposed that the
Fourier modes of square-integrable, bounded mathematical functions can be
used to construct a measure of what they called configurational entropy
(CE): a configuration consisting of a single mode has zero CE (a single wave
in space), while one where all modes contribute with equal weight has
maximal CE. To apply such ideas to physical models, GS used the energy
density of a given spatially-localized field configuration, found from the
solution--exact or approximate--of the related PDE. Of importance in what
follows, GS pointed out that the configurational entropy can be used to
choose the best-fitting trial function in situations where their energies
are degenerate. More generally, the approach presented in \cite%
{PLBgleiser-stamatopoulos} has been recently used to study the
nonequilibrium dynamics of spontaneous symmetry breaking \cite%
{PRDgleiser-stamatopoulos}, to obtain a stability bound for compact
astrophysical objects \cite{PLBgleiser-sowinski}, and to investigate the
emergence of localized objects during inflationary preheating \cite%
{PRDgleiser-graham}.

In the present work we will compute the configurational entropy of some
classes of models with two interacting scalar fields \cite{rajaraman3,
Bazeia, PLBdutra, boya}. These models admit a variety of kink-like
solutions, and have been shown to give rise to bags, junctions, and networks
of BPS and non-BPS defects \cite{Bazeia2}. In particular, we will explore
analytical solutions that are energy-degenerate but quite distinct in their
spatial profiles. We will show that the CE can be used to distinguish
between such configurations, adding a new information-theoretic perspective
to the study of BPS solitons and other nonlinear localized configurations.

Section II introduces the model and its analytical solutions. Section III
reviews the configurational entropy measure for spatially localized
solutions. In section IV we compute the configurational entropy for
two-field BPS solitons and show how it can be used to distinguish between
energy-degenerate configurations. In section V we present our conclusions
and final remarks.

\section{2. Interacting scalar field model and its solutions}

Consider a (1+1)-dimensional model with two interacting real scalar fields
described by the following Lagrangian density%
\begin{equation}
\mathcal{L}=\frac{1}{2}(\partial _{\nu }\phi )^{2}+\frac{1}{2}(\partial
_{\nu }\chi )^{2}-V(\phi ,\chi ),  \label{1}
\end{equation}

\noindent where $V(\phi ,\chi )$ is the potential. We use units with $%
c=\hbar=1$ and metric $\eta _{\nu \beta }=$ diag$(1,-1)$ with coordinates $%
x^{\nu }=(t,x)$.

The potential $V(\phi ,\chi )$ can be represented in terms of a
superpotential $W(\phi ,\chi )$ as%
\begin{equation}
V(\phi ,\chi )=\frac{1}{2}\left[ \left( \frac{\partial W(\phi ,\chi )}{%
\partial \phi }\right) ^{2}+\left( \frac{\partial W(\phi ,\chi )}{\partial
\chi }\right) ^{2}\right] .  \label{2}
\end{equation}

This representation includes supersymmetric models that generate distinct
domain walls and topological solitons \cite{voloshin}-\cite{voloshifman}.

From the Lagrangian density (\ref{1}) and the definition of the
superpotential (\ref{2}), the classical Euler--Lagrange equations of the
static field configurations $\phi =\phi (x)$ and $\chi =\chi (x)$ are given
by%
\begin{eqnarray}
\frac{d^{2}\phi }{dx^{2}} &=&W_{\phi }W_{\phi \phi }+W_{\chi }W_{\chi \phi },
\\
&&  \notag \\
\frac{d^{2}\chi }{dx^{2}} &=&W_{\chi }W_{\chi \chi }+W_{\phi }W_{\chi \phi },
\end{eqnarray}
\noindent where the subscripts denote derivatives with respect to the two
fields. The energy functional of the static field configurations can be
calculated as%
\begin{equation}
E_{BPS}=\frac{1}{2}\int_{-\infty }^{\infty }dx\left[ \left( \frac{d\phi }{dx}%
\right) ^{2}+\left( \frac{d\chi }{dx}\right) ^{2}+W_{\phi }^{2}+W_{\chi }^{2}%
\right] ,  \label{2.1}
\end{equation}

\noindent where $W_{\phi }\equiv \frac{\partial W(\phi ,\chi )}{\partial
\phi }$ and $W_{\chi }\equiv \frac{\partial W(\phi ,\chi )}{\partial \chi }$%
. The above functional energy can be easily rewritten in the following form 
\begin{eqnarray}
E_{BPS} &=&\frac{1}{2}\int_{-\infty }^{\infty }dx\left[ \left( \frac{d\phi }{%
dx}-W_{\phi }\right) ^{2}+\left( \frac{d\chi }{dx}-W_{\chi }\right)
^{2}\right.  \notag \\
&&  \label{2.2} \\
&&\left. +2\left( W_{\phi }\frac{d\phi }{dx}+W_{\chi }\frac{d\chi }{dx}%
\right) \right] .  \notag
\end{eqnarray}

As a consequence, the solutions with minimal energy of the second-order
differential equations for the static solutions can be found from the
following two first-order equations

\begin{equation}
\frac{d\phi }{dx}=W_{\phi },~~\mathrm{and}~~\frac{d\chi }{dx}=W_{\chi }.
\label{3}
\end{equation}

The energy $E_{BPS}$, which is called BPS energy, is written as%
\begin{equation}
E_{BPS}=\left\vert W(\phi _{j},\chi _{j})-W(\phi _{i},\chi _{i})\right\vert ,
\label{3.1}
\end{equation}

\noindent where $\phi _{i}$ and $\chi _{i}$ denote the $i$th vacuum state of
the model.

Following Ref. \cite{PLBdutra}, it is possible from (\ref{3}) to formally
write the equation%
\begin{equation}
\frac{d\phi }{W_{\phi }}=dx=\frac{d\chi }{W_{\chi }},  \label{3.2}
\end{equation}

\noindent which leads to%
\begin{equation}
\frac{d\phi }{d\chi }=\frac{W_{\phi }}{W_{\chi }}.  \label{3.3}
\end{equation}

The above equation is a nonlinear differential equation relating the scalar
fields of the model so that $\phi =\phi (\chi )$. Once this function is
known, equations (\ref{3}) become uncoupled and can be solved.

Considering the application below, we now review the model studied in Refs. 
\cite{Bazeia, Bazeia2, PLBdutra}, used for modeling a great number of
systems \cite{Bazeia2}-\cite{racc2}, whose superpotential is given by

\begin{equation}
W(\phi ,\chi )=-\lambda \phi +\frac{\lambda }{3}\phi ^{3}+\mu \phi \chi ^{2},
\label{4}
\end{equation}

\noindent where $\lambda $ and $\mu $ are real and positive dimensionless
coupling constants. The potential $V(\phi ,\chi )$ of the model with the
above superpotential is given by%
\begin{eqnarray}
&&\left. V(\phi ,\chi )=\frac{1}{2}\left[ \lambda ^{2}+\lambda ^{2}\phi
^{2}(\phi ^{2}-2)\right. \right.  \notag \\
&&  \label{4.1} \\
&&\left. +\mu ^{2}\chi ^{2}\left( \chi ^{2}-\frac{2\lambda }{\mu }\right)
+2\mu ^{2}\left( \frac{\lambda }{\mu }+2\right) \phi ^{2}\chi ^{2}\right] . 
\notag
\end{eqnarray}

For $\lambda /\mu >0$ the model has four supersymmetric minima $(\phi ,\chi
) $%
\begin{eqnarray}
\mathcal{M}_{1} &=&(-1,0),\text{ \ \ }\mathcal{M}_{2}=(1,0),  \notag \\
&&  \label{4.2} \\
\mathcal{M}_{3} &=&\left( 0,-\sqrt{\frac{\lambda }{\mu }}\right) ,\text{ \ \ 
}\mathcal{M}_{4}=\left( 0,\sqrt{\frac{\lambda }{\mu }}\right) .  \notag
\end{eqnarray}

The orbits connecting the vacuum states can be seen on Fig. 1. Note that we
can have six configurations connecting the vacua, where five are BPS states
and one is non-BPS.

\begin{figure}[h]
\begin{center}
\includegraphics[width=8.9cm]{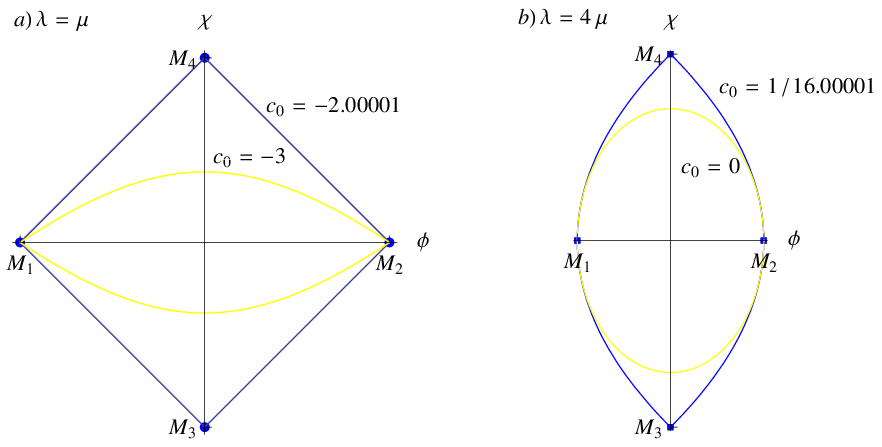} %
\includegraphics[width=8.9cm]{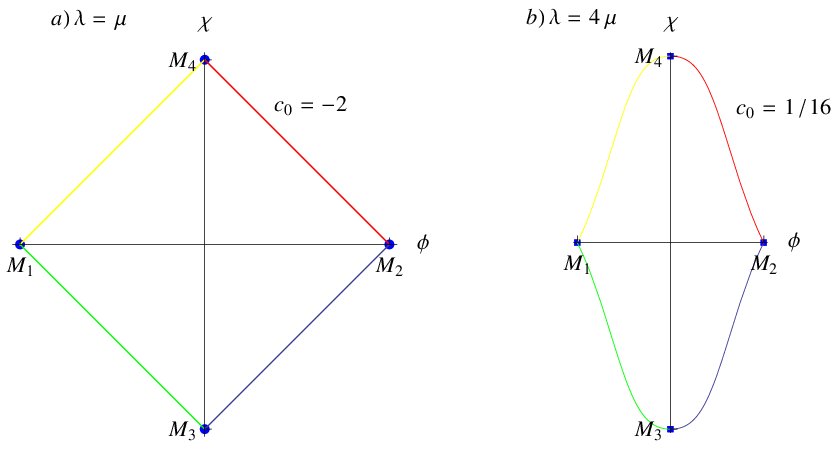}
\end{center}
\caption{Orbit for the solutions and vacuum states of the potential. The
plots on the top of the Figure show the degenerate solutions and the bottom
ones show the critical solutions.}
\label{fig1:vacuum states}
\end{figure}

Using the above results, the sectors connecting the vacua and their
corresponding energies are given by%
\begin{eqnarray}
\mathcal{M}_{1} &\rightarrow &\mathcal{M}_{2},\text{ \ }E_{BPS}^{(12)}=\frac{%
4\lambda }{3},~  \notag \\
\mathcal{M}_{1} &\rightarrow &\mathcal{M}_{3},\text{ \ }E_{BPS}^{(13)}=\frac{%
2\lambda }{3},  \notag \\
\mathcal{M}_{1} &\rightarrow &\mathcal{M}_{4},\text{ \ }E_{BPS}^{(14)}=\frac{%
2\lambda }{3},~  \notag \\
\mathcal{M}_{2} &\rightarrow &\mathcal{M}_{3},\text{ \ }E_{BPS}^{(23)}=\frac{%
2\lambda }{3}, \\
\mathcal{M}_{2} &\rightarrow &\mathcal{M}_{4},\text{ \ }E_{BPS}^{(24)}=\frac{%
2\lambda }{3},~  \notag \\
\mathcal{M}_{3} &\rightarrow &\mathcal{M}_{4},\text{ \ }E_{nBPS}^{(34)}=%
\frac{4\lambda }{3}\sqrt{\frac{\lambda }{\mu }}.  \notag
\end{eqnarray}

Thus, we can see that four sectors have degenerate energies.

As remarked in \cite{PLBdutra}, general solutions of the first-order
differential equations can be found for the scalar fields, by first
integrating the relation

\begin{equation}
\frac{d\phi }{d\chi }=\frac{W_{\phi }}{W_{\chi }}=\frac{\lambda (\phi
^{2}-1)+\mu \chi ^{2}}{2\mu \phi \chi },  \label{5}
\end{equation}%
and then by rewriting one of the fields in terms of the other.

Introducing the new variable $\rho =\phi ^{2}-1$, we can rewrite the above
equation as

\begin{equation}
\frac{d\rho }{d\chi }-\frac{\lambda \rho }{\mu \chi }=\chi ,  \label{6}
\end{equation}

\noindent and the corresponding general solutions are

\begin{eqnarray}
\rho (\chi ) &=&\phi ^{2}-1=c_{0}\chi ^{\lambda /\mu }-\frac{\mu }{\lambda
-2\mu }\chi ^{2},\text{ }(\lambda \neq 2\mu )  \label{7} \\
&&  \notag \\
\rho (\chi ) &=&\phi ^{2}-1=\chi ^{2}[\ln (\chi )+c_{1}],\text{ }(\lambda
=2\mu ),  \label{8}
\end{eqnarray}%
where $c_{0}$ and $c_{1}$ are arbitrary integration constants. Substituting
the above solutions in the first-order differential equation for the field $%
\chi $, we have

\begin{eqnarray}
\frac{d\chi }{dr} &=&\pm 2\mu \chi \sqrt{1+c_{0}\chi ^{\lambda /\mu }-\frac{%
\mu }{\lambda -2\mu }\chi ^{2}},(\lambda \neq 2\mu )  \label{9} \\
&&  \notag \\
\frac{d\chi }{dr} &=&\pm 2\mu \chi \sqrt{1+\chi ^{2}[\ln (\chi )+c_{1}]},%
\text{{\ \ \ }}{(}\lambda =2\mu ).  \label{10}
\end{eqnarray}

It has been found in Ref. \cite{PLBdutra} that in four particular cases the
first equation in (\ref{9}) can be solved analytically. Moreover, in order
keep the solutions finite over all space, $c_{0}$ cannot assume values
higher than some critical ones. At the critical values, the field
configuration changes drastically, as we see next.

\subsection{A. Degenerate Bloch Walls}

Dutra and Hott called the first set of solutions of equation (\ref{9})
degenerate Bloch walls \cite{PRDdutra-hot-amaro} (DBW). There are two
situations with exact classical solutions:

\subsubsection{\textbf{A1. For }$c_{0}<-2$\textbf{\ and }$\protect\lambda =%
\protect\mu $}

In this case we have

\begin{eqnarray}
\chi _{DBW}^{(1)}(x) &=&\frac{2}{\left( \sqrt{c_{0}^{2}-4}\right) \cosh
(2\mu x)-c_{0}},  \label{11} \\
&&  \notag \\
\phi _{DBW}^{(1)}(x) &=&\frac{\left( \sqrt{c_{0}^{2}-4}\right) \sinh (2\mu x)%
}{\left( \sqrt{c_{0}^{2}-4}\right) \cosh (2\mu x)-c_{0}}.  \label{12}
\end{eqnarray}

\subsubsection{\textbf{A2. For }$\protect\lambda =4\protect\mu $\textbf{\
and }$c_{0}<1/16$}

The solutions can be written as

\begin{eqnarray}
\chi _{DBW}^{(2)}(x) &=&-\frac{2}{\sqrt{\left( \sqrt{1-16c_{0}}\right) \cosh
(4\mu x)+1}},  \label{13} \\
&&  \notag \\
\phi _{DBW}^{(2)}(x) &=&\frac{\left( \sqrt{1-16c_{0}}\right) \sinh (4\mu x)}{%
\left( \sqrt{1-16c_{0}}\right) \cosh (4\mu x)+1}.  \label{14}
\end{eqnarray}

\begin{figure}[h]
\begin{center}
\includegraphics[width=8.8cm]{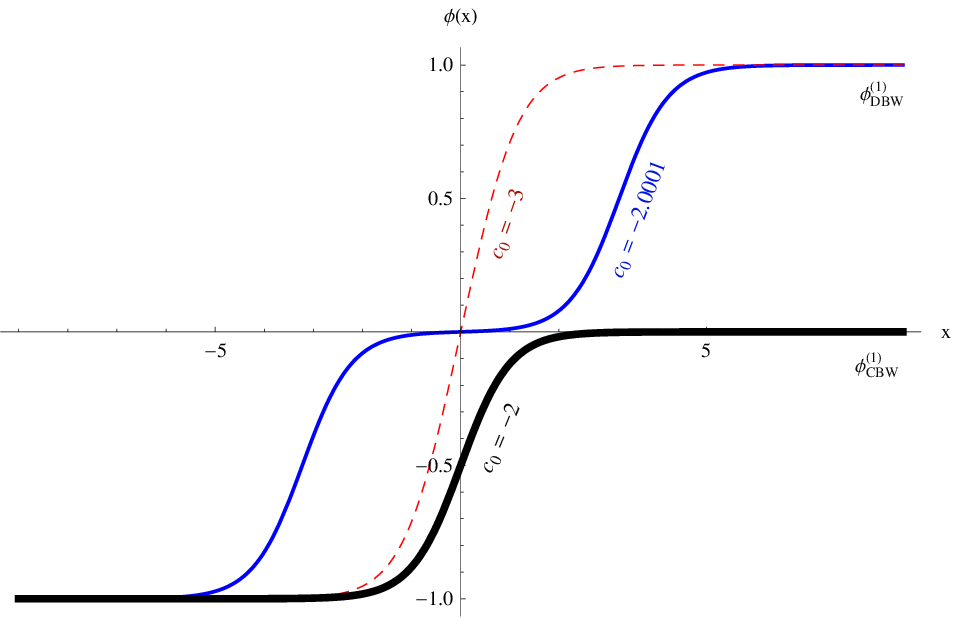} \\[1cm]
\includegraphics[width=8.8cm]{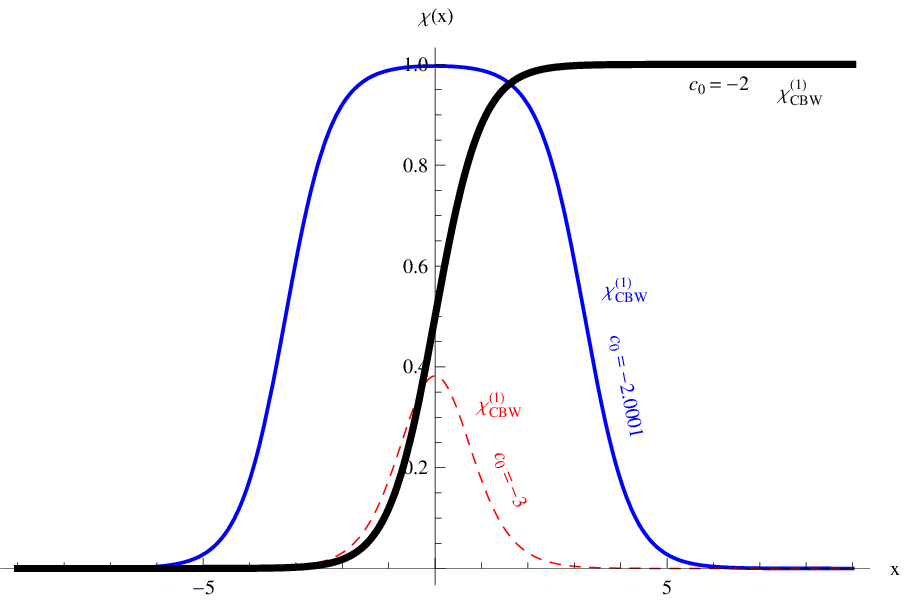}
\end{center}
\caption{Sample profiles of the DBW and CBW solutions. Thin lines
(continuous and dashed) are for two DBW solutions, while the thick line is
for the CBW solution.}
\label{fig2:DBW1}
\end{figure}

In Figure \ref{fig2:DBW1} we show some typical profiles of the DBW
solutions. Note that the two-kink solution (top) $\phi _{\mathrm{DBW}}^{(1)}$
arises only for values of $c_{0}$ close to the critical value, $c_{0}^{(%
\mathrm{crit)}}=-2$. For the same values of $c_0$, the corresponding
lump-like solutions for $\chi _{\mathrm{DBW}}^{(1)}$(bottom) exhibit a flat
top, which disappears as we move away from $c_{0}^{(\mathrm{crit)}}$. As we
will see, the related CE for these configurations carry a very distinctive
signature.

An important feature of the DBW solutions is that their energies are
degenerate with respect to $c_{0}$: for any value of $c_{0}$ the energy is
given by $E_{DBW}=4\lambda /3$. This means that energy alone cannot
distinguish between the rich variation in the spatial profiles of the DBW
solutions as $c_0$ is varied. As we shall see, this is where the CE will
play a key role.

\subsection{B. Critical Bloch Walls}

An interesting class of analytical solutions, named as critical Bloch walls
(CBW), was shown to exist when the constant of integration is taken to be
equal to the critical value. Again, we have two cases:

\subsubsection{\textbf{B1. For }$\protect\lambda =\protect\mu $\textbf{\ and 
}$c_{0}=-2$}

One has the following set of solutions for the scalar fields

\begin{eqnarray}
\chi _{CBW}^{(1)}(x) &=&\frac{1}{2}\left[ 1\pm \tanh (\mu x)\right] ,
\label{15} \\
&&  \notag \\
\phi _{CBW}^{(1)}(x) &=&-\frac{1}{2}\left[ \tanh [\mu x)\mp 1\right] .
\label{16}
\end{eqnarray}

\subsubsection{\textbf{B2. For }$\protect\lambda =4\protect\mu ~$\textbf{and 
}$c_{0}=1/16$}

Now, the solutions for the fields are given by

\begin{eqnarray}
\chi _{CBW}^{(2)}(x) &=&\sqrt{2}{\ }\frac{\cosh (\mu x)\pm \sinh (\mu x)}{%
\sqrt{\cosh (2\mu x)}},  \label{17} \\
&&  \notag \\
\phi _{CBW}^{(2)}(x) &=&\frac{1}{2}\left[ \pm 1-\tanh (2\mu x)\right] .
\label{18}
\end{eqnarray}

In Fig. \ref{fig2:DBW1} we also show the case CBW for $c_{0}-2$. Now, the
energy is $E_{CBW}=2\lambda /3$, consistent with the energy for the DBW case
since one can use two CBW configurations in order to connect the vacua
connected by the DBW: $E_{DBW}=4\lambda /3=2~E_{CBW}$, as can be seen from
Fig. 1.

\section{3. Configurational entropy for two interacting scalar fields}

Recently, GS showed that a configurational entropy measure in functional
space can be used to discriminate between same-energy spatially-localized
solutions \cite{PLBgleiser-stamatopoulos}. The configurational entropy (CE)
is defined as%
\begin{equation}
S_{c}[f]=-\int d^{d}\mathbf{k}\tilde{f}(\mathbf{k})\ln [\tilde{f}(\mathbf{k}%
)],  \label{19}
\end{equation}

\noindent where $d$ is the number of spatial dimensions, and $\tilde{f}(%
\mathbf{k})=f(\mathbf{k})/f_{\max }(\mathbf{k})$. The function $f(\mathbf{k})
$ is defined as the modal fraction%
\begin{equation}
f(\mathbf{k})=\frac{\left\vert F(\mathbf{k})\right\vert ^{2}}{\int d^{d}%
\mathbf{k}\left\vert F(\mathbf{k})\right\vert ^{2}}.  \label{20}
\end{equation}

\noindent $f_{\max }(\mathbf{k})$ is the maximal modal fraction, that is,
the mode giving the highest contribution. This normalization guarantees that 
$S_c[f]$ is positive-definite. The function $F(\mathbf{k})$ represents the
Fourier transform of the energy density of the configuration. In order to
compute the CE the energy density must be square-integrable even if the
fields are not.

Here, we will extend the procedure presented in \cite%
{PLBgleiser-stamatopoulos} to models with two coupled fields. For static
configurations of two interacting real scalar fields the energy density is
written as%
\begin{equation}
\rho (x)=\frac{1}{2}\left[ \left( \partial _{x}\phi \right) ^{2}+\left(
\partial _{x}\chi \right) ^{2}+V(\phi ,\chi )\right] .
\end{equation}

Following the approach presented in \cite{PLBgleiser-stamatopoulos}, we use
the energy density of the DBW and CBW configurations to compute their
related CE. The Fourier transform is given by%
\begin{equation}
F(k)=\frac{1}{\sqrt{2\pi }}\int_{-\infty }^{\infty }dxe^{ikx}\rho (x),
\label{1.1.1}
\end{equation}

Plancherel's theorem states that%
\begin{equation}
\int_{-\infty }^{\infty }dx\left\vert \rho (x)\right\vert ^{2}=\int_{-\infty
}^{\infty }dk\left\vert F(k)\right\vert ^{2}.  \label{1.1.3}
\end{equation}

Again, we stress that the spatially-localized energy densities must be
square-integrable bounded functions $\rho (x)\in L^{2}(\mathbf{R})$.

\section{4. Entropy for DBW and CBW}

We now use the approach presented in the previous section to obtain the
configurational entropy of the DBW and the CBW configurations. Let us begin
with the DBW case, which has two sets of exact solutions, given in section
II.A. Starting with case A1, $\lambda =\mu $ and $c_{0}<-2$, from the scalar
fields given in equations (\ref{11}) and (\ref{12}), we obtain the
corresponding energy density as%
\begin{eqnarray}
&&\left. \rho _{DBW}^{(1)}(x)=\frac{6\mu ^{2}}{\left[ a_{1}+\cosh (2\mu x)%
\right] ^{4}}\right.  \notag \\
&&  \label{22.1.2} \\
&&\left. -\frac{8\mu ^{2}c_{0}\cosh (2\mu x)}{\alpha \left[ a_{1}+\cosh
(2\mu x)\right] ^{4}}+\frac{2\mu ^{2}\cosh (4\mu x)}{\left[ a_{1}+\cosh
(2\mu x)\right] ^{4}},\right.  \notag
\end{eqnarray}

\noindent where $\alpha =\alpha (c_{0})\equiv c_{0}/\sqrt{c_{0}^{2}-4}$ and $%
a_{1}\equiv -c_{0}/\alpha $.

On the other hand, for case A2, where $\lambda =4\mu $ and $c_{0}<1/16$, the
corresponding energy density is%
\begin{eqnarray}
&&\left. \rho _{DBW}^{(2)}(x)=-\frac{16\mu ^{2}[\beta ^{2}+\cosh (2\mu x)]}{%
\beta ^{2}[a_{2}+\cosh (4\mu x)]^{4}}\right.  \notag \\
&&  \label{23.1.1} \\
&&\left. -\frac{4\mu ^{2}[7\cosh (4\mu x)+\cosh (12\mu x)]}{\beta \lbrack
a_{2}+\cosh (4\mu x)]^{4}},\right.  \notag
\end{eqnarray}

\noindent with $\beta =\beta (c_{0})\equiv \sqrt{1-16c_{0}}$ and $%
a_{2}\equiv 1/\beta $.

At this point, we note again that different solutions to the DBW equations
have energies that are degenerate with respect to $c_{0}$. In \cite%
{PLBgleiser-stamatopoulos}, Gleiser and Stamatopoulos studied a case where
different trial functions used to approximate the actual solution were
energy-degenerate. They showed that the CE could be used to select which of
the trial functions was a better fit to the exact solution: that which had
minimal CE. Here, we have a novel situation where the actual analytical
solutions to the equations of motion have an infinite degeneracy with
respect to a single parameter ($c_0$). We will follow the approach in GS and
examine whether the CE can be used to discriminate between solutions which
are energy-degenerate. In this way, we are proposing that the
configurational entropy is an excellent tool to resolve ambiguous situations
that may emerge from Hamilton's variational principle.

We thus proceed to compute the Fourier transform of the energy density (\ref%
{22.1.2}) and (\ref{23.1.1}), which gives the modal fraction (\ref{20}).
Using (\ref{1.1.1}), we have%
\begin{eqnarray}
F^{(1)}(k) &=&\frac{1}{\sqrt{2\pi }}\int_{-\infty }^{\infty }dxe^{ikx}\rho
_{DBW}^{(1)}(x),  \label{f1.1} \\
&&  \notag \\
F^{(2)}(k) &=&\frac{1}{\sqrt{2\pi }}\int_{-\infty }^{\infty }dxe^{ikx}\rho
_{DBW}^{(2)}(x).  \label{f2.1}
\end{eqnarray}

In order to obtain an analytical expression for the above Fourier
transforms, it is useful to introduce the generalized integral%
\begin{equation}
I^{(n)}(a_{n},\gamma ,\delta ,k)=\int_{-\infty }^{\infty }dx\frac{%
e^{ikx}\cosh (\gamma x)}{[a_{n}+\cosh (\delta x)]^{4}}.  \label{i.1}
\end{equation}

After lengthy but straightforward calculations, one finds%
\begin{equation}
I^{(n)}(a^{n},\gamma ,\delta ,k)=\frac{8}{\delta }\sum%
\limits_{j=1}^{2}G_{j}(a_{n},\gamma ,\delta ,k),  \label{i.2}
\end{equation}

\noindent where%
\begin{eqnarray}
&&\left. G_{j}(a_{n},\gamma ,\delta ,k)=\frac{1}{\Omega _{j}+4}%
F_{1}[A_{j};B_{j},B_{j}^{\prime };C_{j};X_{n},Y_{n}]\right.  \notag \\
&&  \label{i.3} \\
&&-\frac{1}{\Omega _{j}-4}F_{1}[\bar{A}_{j};\bar{B}_{j},\bar{B}_{j}^{\prime
};\bar{C}_{j};X_{n},Y_{n}],  \notag
\end{eqnarray}

\noindent and the functions $F_{1}[A_{j};B_{j},B_{j}^{\prime
};C_{j};X_{n},Y_{n}]$ and $F_{1}[\bar{A}_{j};\bar{B}_{j},\bar{B}_{j}^{\prime
};\bar{C}_{j};X_{n},Y_{n}]$ are the so-called Appell hypergeometric
functions of two variables with 
\begin{eqnarray*}
&&\left. \Omega _{j}=i\omega +(-1)^{j+1}\Omega ,\text{ }\omega =k/\delta ,%
\text{ }\Omega =\gamma /\delta ,\right. \\
&& \\
&&\left. A_{j}=\Omega _{j}+4,\text{ }B_{j}=B_{j}^{\prime }=4,\text{ }%
C_{j}=\Omega _{j}+5,\right. \\
&& \\
&&\left. \bar{A}_{j}=-\Omega _{j}+4,\text{ }\bar{B}_{j}=\bar{B}_{j}^{\prime
}=4,\text{ }\bar{C}_{j}=-\Omega _{j}+5,\right. \\
&& \\
&&\left. X_{n}=-1/\left[ a_{n}-\sqrt{a_{n}^{2}-1}\right] ,\right. \\
&& \\
&&\left. Y_{n}=-1/\left[ a_{n}+\sqrt{a_{n}^{2}-1}\right] .\right.
\end{eqnarray*}

We can now write the Fourier transforms of (\ref{f1.1}) and (\ref{f2.1}) in
the following compact forms%
\begin{eqnarray}
&&\left. F^{(1)}(k)=\frac{2\mu ^{2}}{\sqrt{2\pi }}\left[ 3I^{(1)}(a_{1},0,2%
\mu ,k)\right. \right.  \notag \\
&&  \notag \\
&&\left. -\frac{4c_{0}}{\alpha }I^{(1)}(a_{1},2\mu ,2\mu
,k)+I^{(1)}(a_{1},4\mu ,2\mu ,k)\right] , \\
&&  \notag \\
&&\left. F^{(2)}(k)=-\frac{4\mu ^{2}}{\sqrt{2\pi }}\left[ I^{(2)}(a_{2},0,4%
\mu ,k)\right. \right.  \notag \\
&&  \notag \\
&&\left. +\frac{4}{\beta ^{2}}I^{(2)}(a_{2},2\mu ,4\mu ,k)+\frac{7}{\beta }%
I^{(2)}(a_{2},4\mu ,4\mu ,k)\right.  \notag \\
&&  \notag \\
&&\left. \left. +\frac{1}{\beta }I^{(2)}(a_{2},12\mu ,4\mu ,k)\right]
.\right.
\end{eqnarray}

In order to obtain the modal fraction (\ref{20}) it is necessary to evaluate
(\ref{1.1.3}), clearly a daunting task. To proceed analytically, we evaluate
the integrals for $F^{(1)}(k)$ and $F^{(2)}(k)$ numerically, and fit them as
functions of the single parameter $c_0$ as 
\begin{eqnarray}
\int_{-\infty }^{\infty }dk\left\vert F^{(1)}(k)\right\vert ^{2} &\simeq
&g_{1}-g_{2}e^{g_{3}c_{0}}, \\
&&  \notag \\
\int_{-\infty }^{\infty }dk\left\vert F^{(2)}(k)\right\vert ^{2} &\simeq
&h_{1}-h_{2}e^{h_{3}c_{0}},
\end{eqnarray}

\noindent where $g_{1}=0.8481$, $g_{2}=3.8834$, $g_{3}=1.1332$, $%
h_{1}=41.0711$, $h_{2}=23.2854$ and $h_{3}=1.1699$.

The modal fractions can be approximated by%
\begin{equation}
f^{(1)}(k)\simeq \frac{\left\vert F^{(1)}(k)\right\vert ^{2}}{%
g_{1}-g_{2}e^{g_{3}c_{0}}},~\text{ }f^{(2)}(k)\simeq \frac{\left\vert
F^{(2)}(k)\right\vert ^{2}}{h_{1}-h_{2}e^{h_{3}c_{0}}}.
\end{equation}

These modal fractions are plotted in Fig. \ref{fig3:modal fractions}. As can
be seen, they are localized and exhibit a maximum at $k=0$. These
expressions are to be used into equation (\ref{19}) in order to obtain the
CE for each of the two DBW cases:

\begin{figure}[h]
\begin{center}
\includegraphics[width=8.6cm]{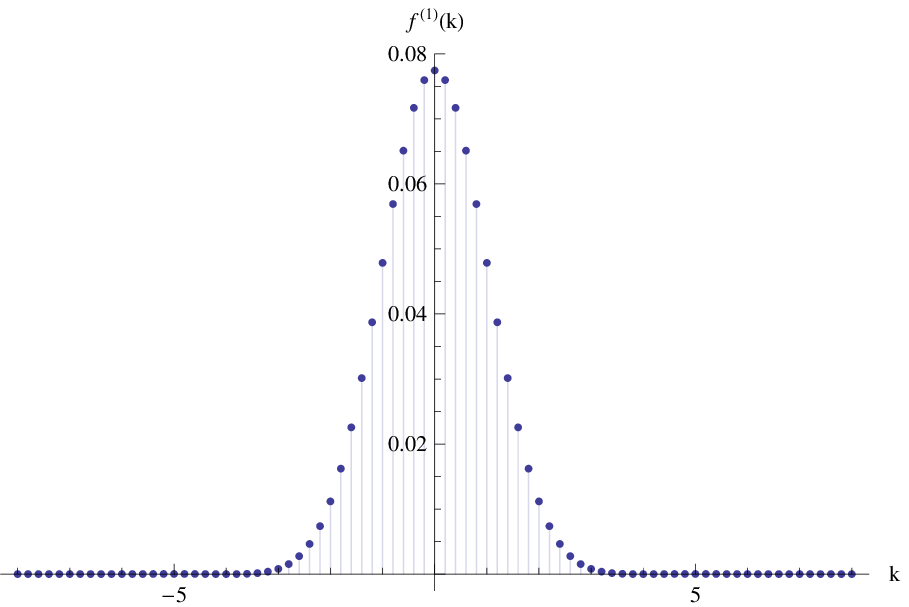} \\[0.8cm]
\includegraphics[width=8.6cm]{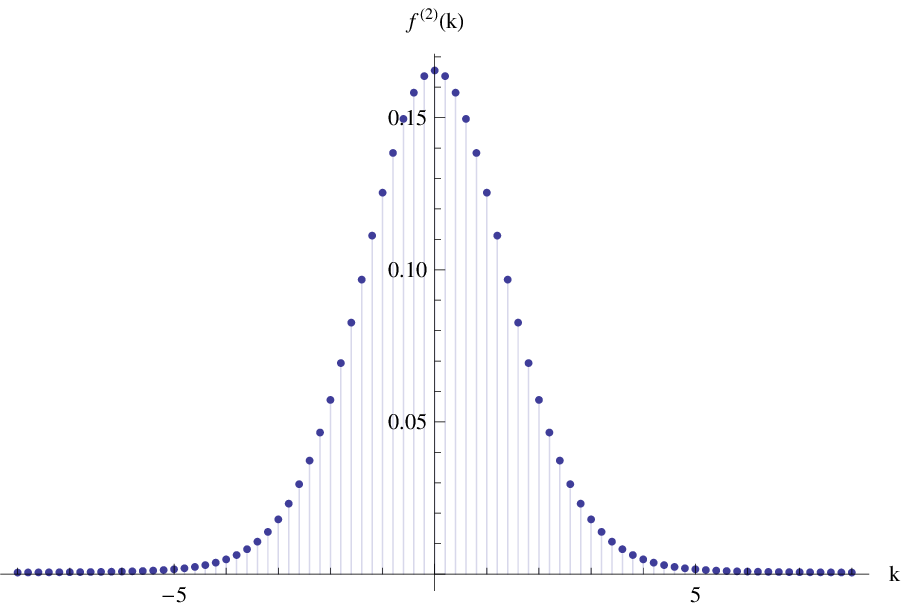}
\end{center}
\caption{Modal fractions with $\protect\mu=1$. Note that the maximum is at $%
k=0$.}
\label{fig3:modal fractions}
\end{figure}

\begin{eqnarray}
S_{c}^{(1)} &\simeq &-\int dk\tilde{f}^{(1)}(k)\ln [\tilde{f}^{(1)}(k)], \\
&&  \notag \\
S_{c}^{(2)} &\simeq &-\int dk\tilde{f}^{(2)}(k)\ln [\tilde{f}^{(2)}(k)].
\label{CEnum}
\end{eqnarray}

\begin{figure}[h]
\begin{center}
\includegraphics[width=8.9cm,height=8.5cm]{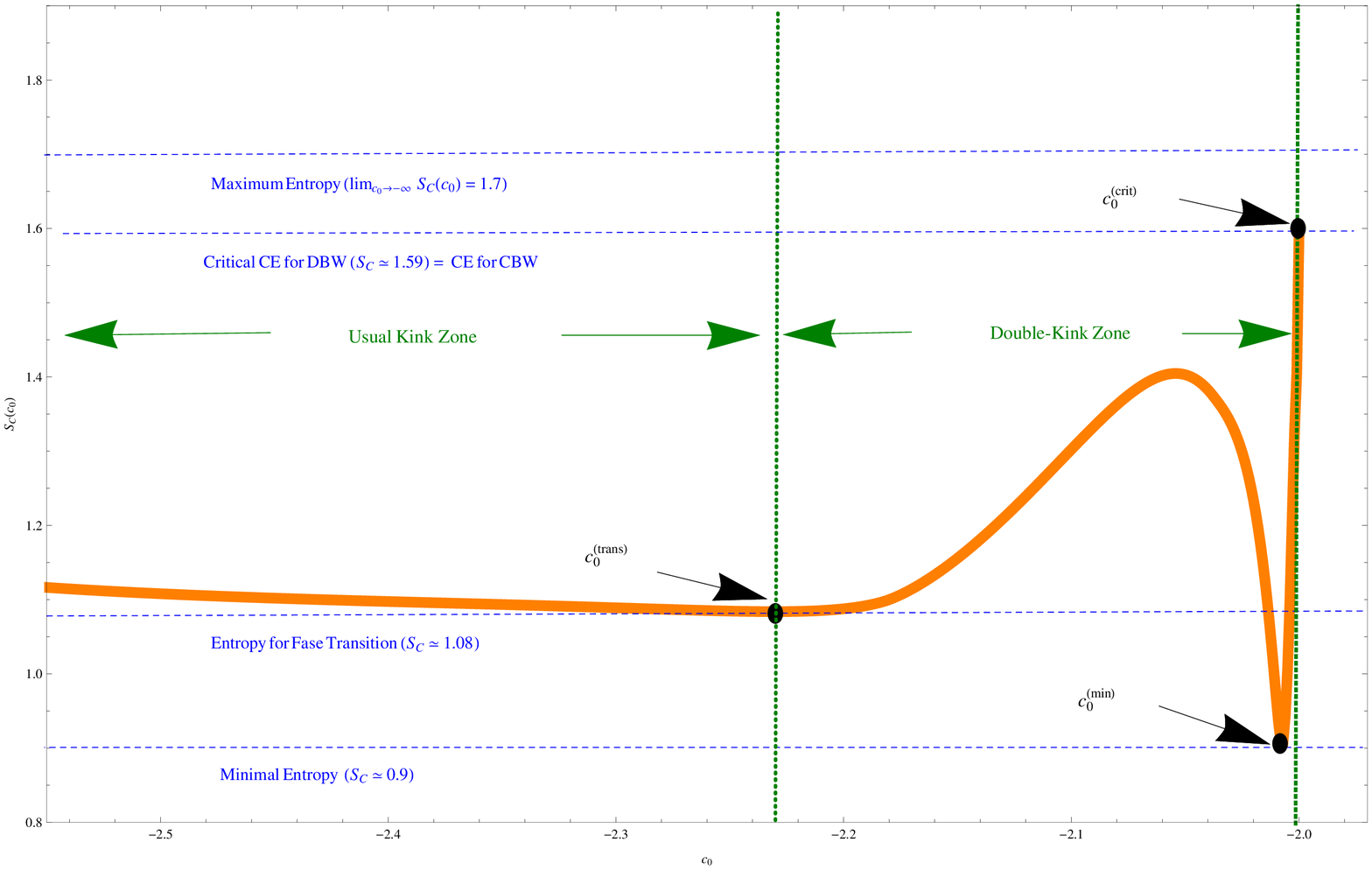} \\[1cm]
\includegraphics[width=8.9cm,height=8.5cm]{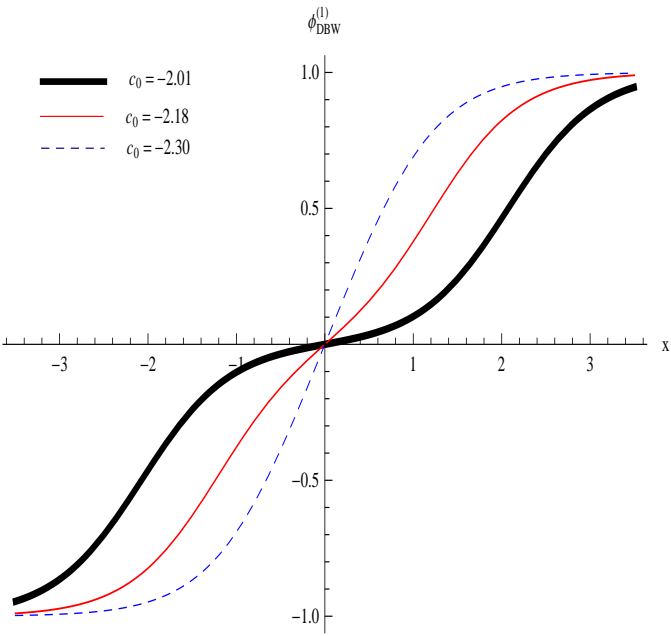}
\end{center}
\caption{The configurational entropy (top) for DBW solutions as a function
of the parameter $c_0$ and a few of its corresponding $\protect\phi _{%
\mathrm{DBW}}^{(1)}$-field configurations (bottom). The curves correspond to
choosing $\protect\lambda =1$.}
\label{fig4:Entropy}
\end{figure}

To compute the configurational entropy, we must integrate equations (\ref%
{CEnum}) numerically. The results are shown in Fig. \ref{fig4:Entropy},
where the CE is plotted as a function of the parameter $c_0$. It is quite
remarkable that the CE shows such rich structure for varying $c_0$ while the
energies for all these configurations are simply degenerate. There is a
sharp minimum at the value $c_0^{(\mathrm{min})} \simeq -2.005$, the region
of parameter space where the double-kink solution is most prominent, within
our numerical accuracy. This can be seen by plotting the three inflection
points of the solution for $\phi _{\mathrm{DBW}}^{(1)}$, and showing how
they progressively merge into a single inflection point--a single kink--as $%
c_0$ is decreased. Below $c_{0}^{(\mathrm{trans)}}$, an inflection point for
CE, the field configurations undergo a quick transition, where the two-kink
solution in the field $\phi _{\mathrm{DBW}}^{(1)}$ rapidly converges into a
single kink at $c_0\lesssim -2.30$, while the lump-like solutions for $\chi
_{\mathrm{DBW}}^{(1)}$, which have a flat-top profile for $c_{0}> c_{0}^{(%
\mathrm{trans)}}$, become more Gaussian-like.

For completeness, we examined how the results vary with respect to the
two-field coupling constant $\lambda$, taken to be $\lambda=1$ in Fig. \ref%
{fig4:Entropy}. The qualitative features remain the same, at least for these
values of $\lambda$. The results are shown in Table \ref{Table 1}:

\begin{table}[h]
\begin{center}
\includegraphics[width=8.8cm,height=2.5cm]{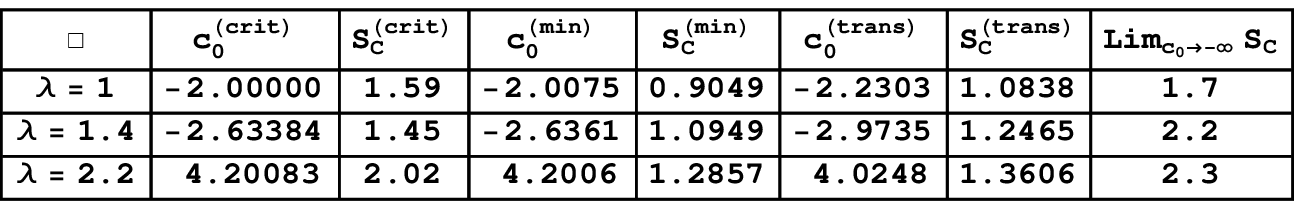}
\end{center}
\caption{Configurational entropy and values of corresponding $c_{0}$ for
various choices of the two-field coupling $\protect\lambda$ for Degenerate
Bloch Walls.}
\label{Table 1}
\end{table}

\section{5. Conclusions.}

We investigated the properties of kink-like solutions in a class of
interacting two-field models in two spacetime dimensions. The models we
considered are especially interesting because we can use the BPS approach to
find exact analytical solutions. In particular, a class of these solutions
know as Degenerate Bloch Walls is degenerate in energy, even if their
spatial profiles are widely different. Using the configurational entropy
measure of Gleiser and Stamatopoulos, we were able to extract information
about the different solutions which is clearly related to their spatial
profiles. We thus propose that this information-entropic measure is an
essential tool in the study of complex spatially-localized configurations,
providing valuable information beyond simple energetics.

\bigskip

\textbf{Acknowledgements}

The authors A. de Souza Dutra and R. A. C. C. are grateful to CNPq and CAPES
for partial financial support. R. A. C. C. thanks to the Professor Denis
Dalmazi for discussions on questions concerning statistical mechanics and
Dartmouth College for partial finantial support. MG was supported in part by
a National Science Foundation grant PHY-1068027 and by a Department of
Energy grant DE-SC0010386. MG also acknowledges support from the John
Templeton Foundation grant under the New Frontiers in Astronomy \& Cosmology
program and under grant no. 48038.


\bigskip

\begin{thebibliography}{99}
\bibitem{Russell} J. S. Russell, \textit{Report on Waves}, Fourteenth
meeting of the British Association for the Advancement of Science, \textbf{14%
} (1844) 311.

\bibitem{Whitham} G. B. Whitham, \textit{Linear and Non-Linear Waves}, John
Wiley and Sons, New York, (1974).

\bibitem{Scott} A. C. Scott, F. Y. F. Chiu and D. W. Mclaughlin, Proc.
I.E.E.E. \textbf{61} (1973) 1443.

\bibitem{Jackiw} R. Jackiw and C. Rebbi, Phys. Rev. D \textbf{13} (1975)
3398.

\bibitem{Lee} R. Friedberg, T. D. Lee, and A. Sirlin, Phys. Rev. D \textbf{13%
} (1976) 2739.

\bibitem{Makhankov} V. G. Makhankov, Phys. Reports \textbf{35 }(1978) 1.

\bibitem{Rajaraman1} R. Rajaraman and E. J. Weinberg, Phys. Rev. D \textbf{11%
} (1975) 2950.

\bibitem{Vilenkin} A. Vilenkin and E. P. S. Shellard, \textit{Cosmic Strings
and Other Topological Defects} (Cambridge University, Cambridge, England,
1994).

\bibitem{Vachaspati} T. Vachaspati, \textit{Kinks and Domain Walls: An
Introduction to Classical and Quantum Solitons} (Cambridge University Press,
Cambridge, England, 2006).

\bibitem{Weinberg} E. J. Weinberg, \textit{Classical Solutions in Quantum
Field Theory: Solitons and Instantons in High Energy Physics }(Cambridge
University Press, Cambridge, England, 2012).

\bibitem{Bishop} A. R. Bishop and T. Schneider, \textit{Solitons and
Condensed Matter Physics }(Springer-Verlag, Berlin, 1978).

\bibitem{Gu} C. Gu, \textit{Soliton Theory and Its Applications }%
(Springer-Verlag, Berlin, 1995).

\bibitem{Rajaraman2} R. Rajaraman, \textit{Solitons and Instantons}
(North-Holand, Amsterdam, 1982).

\bibitem{Dashen} R. F. Dashen, B. Hasslacher, and A. Neveu, Phys. Rev. D 
\textbf{10} (1974) 4130.

\bibitem{Bogomolnyi} E. B. Bogomolnyi, Sov. J. Nucl. Phys. \textbf{24}
(1976) 449.

\bibitem{Prasad} M. K. Prasad and C. M. Sommerfield, Phys. Rev. Lett. 
\textbf{35}, (1975) 760.

\bibitem{Adam} C. Adam, J. M. Queiruga, J. Sanchez-Guillen, and A.
Wereszczynski, J. High Energy Phys. \textbf{1305} (2013) 108.

\bibitem{laf} L. A. Ferreira and Wojtek J. Zakrzewski, J. High Energy Phys. 
\textbf{1309} (2013) 097.

\bibitem{Bobby} B. Cheng and C. Ford, Phys. Lett. B \textbf{720 }(2013) 262.

\bibitem{Casana} R. Casana, M. M. Ferreira, Jr, and E. da Hora, Phys.Rev. D 
\textbf{86} (2012) 085034.

\bibitem{Halmagyi} N. Halmagyi, M. Petrini, and A. Zaffaroni, J. High Energy
Phys. \textbf{1308} (2013) 124.

\bibitem{Cassani} D. Cassani, G. Dall'Agata, and A. F. Faedo, J. High Energy
Phys. \textbf{1303} (2013) 007.

\bibitem{Adam2} C. Adam, J. M. Queiruga, J. Sanchez-Guillen, and A.
Wereszczynski, Phys.Rev. D \textbf{86} (2012) 105009.

\bibitem{rajaraman3} R. Rajaraman, Phys. Rev. Lett. \textbf{42} (1979) 200

\bibitem{Bazeia} D. Bazeia, M. J. dos Santos, andqa R. F. Ribeiro, Phys.
Lett. A \textbf{208} (1995) 84.

\bibitem{PLBdutra} A. de Souza Dutra, Phys. Lett. B \textbf{626 }(2005) 249.

\bibitem{PLBgleiser-stamatopoulos} M. Gleiser and N. Stamatopoulos, Phys.
Lett. B \textbf{713 }(2012) 304.

\bibitem{PRDgleiser-stamatopoulos} M. Gleiser and N. Stamatopoulos, Phys.
Rev. D \textbf{86 }(2012) 045004.

\bibitem{PLBgleiser-sowinski} M. Gleiser and D. Sowinski, Phys. Lett. B 
\textbf{727 }(2013) 272.

\bibitem{PRDgleiser-graham} M. Gleiser and N. Graham, Phys. Rev. D \textbf{89%
} (2014) 083502.

\bibitem{boya} L. J. Boya, J. Casahorran, Phys. Rev. A \textbf{39 }(1989)
4298.

\bibitem{Bazeia2} D. Bazeia, F. A. Brito, Phys. Rev. D. \textbf{61 }(2000)
105019.

\bibitem{voloshin} M. B. Voloshin, Phys. Rev. D. \textbf{57 }(1997) 1266.

\bibitem{shifman} M. Shifman, Phys. Rev. D. \textbf{57 }(1997) 1258.

\bibitem{voloshifman} M. A. Shifman and M. B. Voloshin, Phys. Rev. D. 
\textbf{57 }(1998) 2590.

\bibitem{bazeia3} E. Ventura, A. M. Simas, and D. Bazeia, Chem. Phys. Lett. 
\textbf{320} (2000) 587.

\bibitem{salamanca} A. Alonso Izquierdo, M. A. Gonzalez Leon, J. Mateos
Guilarte, Phys. Rev. D \textbf{65 }(2002) 085012.

\bibitem{bazeia4} M. N. Barreto, D. Bazeia, and R. Menezes, Phys. Rev. D 
\textbf{73 }(2006) 065015.

\bibitem{fabricio} A. de Souza Dutra, M. Hott, and F. A. Barone, Phys. Rev.
D \textbf{74 }(2006) 085030.

\bibitem{racc1} R. A. C. Correa, A de Souza Dutra, and M B Hott, Class.
Quant. Grav. \textbf{28} (2011) 155012.

\bibitem{racc2} A. de Souza Dutra and R. A. C. Correa, Phys. Rev. D \textbf{%
83} (2011) 105007.

\bibitem{PRDdutra-hot-amaro} A. de Souza Dutra, A. C. Amaro de Faria Jr, and
M. Hott, Phys. Rev. D \textbf{78 }(2008) 043526.
\end{thebibliography}
\end{document}